\documentclass[10pt,DIV11]{scrartcl}
\usepackage[latin1]{inputenc}
\usepackage{graphicx}
\usepackage[english]{babel}

\usepackage{amsmath}
\usepackage{amssymb}
\usepackage{amsthm}
\usepackage{amsopn}
\usepackage{pst-all}
\usepackage{subfigure}

\newcommand{\Real}{\mathbb{R}}

\newcommand{\T}{^\mathrm{T}}

\begin{document}

\title{Bridging Time Scales in Cellular Decision Making with a Stochastic Bistable Switch}

\author{
Steffen Waldherr, Jingbo Wu, and Frank Allg{\"o}wer 
}

\date{}

\maketitle

\vspace*{-1.5cm}
\begin{center}
\begin{small}
Institute for Systems Theory and Automatic Control,\\ Universit\"at Stuttgart, Pfaffenwaldring 9,\\ Stuttgart, Germany
\end{small}
\end{center}

\begin{abstract}
Cellular transformations which involve a significant phenotypical change of the cell's state use bistable biochemical switches as underlying decision systems.
      
In this work, we aim at linking cellular decisions taking place on a time scale of years to decades with the biochemical dynamics in signal transduction and gene regulation, occuring on a time scale of minutes to hours.
We show that a stochastic bistable switch forms a viable biochemical mechanism to implement decision processes on long time scales.
As a case study, the mechanism is applied to model the initiation of follicle growth in mammalian ovaries, where the physiological time scale of follicle pool depletion is on the order of the organism's lifespan.
We construct a simple mathematical model for this process based on experimental evidence for the involved genetic mechanisms.

Despite the underlying stochasticity, the proposed mechanism turns out to yield reliable behavior in large populations of cells subject to the considered decision process.
Our model explains how the physiological time constant may emerge from the intrinsic stochasticity of the underlying gene regulatory network.
Apart from ovarian follicles, the proposed mechanism may also be of relevance for other physiological systems where cells take binary decisions over a long time scale.
\end{abstract}

\section*{Background}

The dynamics of biological systems span a wide range of temporal and spatial scales.
The interactions among dynamical properties on different scales govern the overall behavior of the biological system, and thus form an important area of computational research in biology \cite{MartinsFer2009}.
A particularly interesting question in this context is how the behavior on a slow time scale emerges mechanistically from the dynamics on fast time scales.
For example, how do cell population dynamics in tissues, which may evolve on a time scale of months, years or even decades, originate from the dynamics of the underlying gene regulatory networks, with a time scale of just minutes to hours?

In this work, we aim at bridging the time scale from gene regulation to cellular transformation processes on the tissue or cell population level.
We specifically consider cellular transformation processes based on a bistable biochemical switch.
Such switches have two distinct stable stationary states, and the cell initiates a transformation when the switch changes from one stable state to the other one.
Bistable switches have previously been used to model a large number of cellular transformation events, such as progression through cell cycle arrest in the maturation of \textit{Xenopus} oocytes \cite{FerrellXio2001,FerrellMac1998} or initiation of programmed cell death \cite{EissingCon2004} and cellular differentiation \cite{ChickarmaneEnv2009} in higher organisms.
Most models for these systems are constructed as deterministic models, and thus an external stimulus is required to induce changes in the switch's state.

Here, we consider transformation processes which apparently
do not require any external stimulus to be initiated, and which still follow
reliable temporal characteristics.
Reliable thereby means that in a large population of cells, the number of
cells that have already initiated the transformation can be described
deterministically with high accuracy.
We propose a generic transformation process, where a phenotypical change in
the state of a cell is initiated as soon as a bistable biochemical switch
changes its internal state.
In previous studies, random switching caused by internal fluctuations is usually attributed to pathological events \cite{IsaacsHas2003}.
In the mechanism proposed here, random switching has a regular physiological function.

A striking example for the kind of transformation processes we aim
to describe is involved in mammalian oocyte maturation.
In mammalian females, all or almost all of the oocytes that will
ovulate through the organism's life-span are already present at birth or shortly thereafter
as a population of so-called primordial follicles.
Throughout the organism's reproductive life, follicles undergo the
primordial to primary transition, which marks the start of a development
process that will eventually lead to either ovulation or removal of the
oocyte through atresia \cite{FortuneCus2000,Skinner2005}.
In this way, there is a steady supply of mature follicles for ovulation, while
the pool of primordial follicles is gradually depleted.
The mechanisms through which the follicle transition is initiated are largely unknown,
although a number of ovarian factors that may be relevant have been
identified experimentally \cite{ParrottSki1999,CastrillonMia2003,NilssonSki2004}.
Importantly, the transition seems to be regulated locally in the ovary, and
not through the endocrine system \cite{Braw-Tal2002}.
An astonishing observation in this process is that in one follicle,
the transition may occur already
a few months after generation of the primordial follicle pool,
while another follicle may stay several decades (for organisms
with a sufficiently long lifespan)
in the resting stage before growth is initiated.
From the medical side, a misregulation of this process is implicated in premature
ovarian failure due to follicle depletion, which is a major reason for
infertility in human females.

By way of a case study, we apply the proposed transformation mechanism to the problem of growth initiation in ovarian follicles.
Including also cell--cell interactions supported by experimental evidence, we obtain a physiologically plausible model for this process, showing very good agreement with human clinical data on a scale of several decades.

\section*{Models and Methods}
\label{sec:methods}

\subsection*{Deterministic model of a bistable switch}
\label{ssec:switch-model}

The model of a bistable switch that we use is based on a positive feedback
loop between two components.
Consider a biochemical reaction network involving the two molecular species X and Y.
Mathematically, the temporal evolution of the amounts of the two species
is described with the ordinary differential equation
\begin{equation}\begin{aligned}
\dot x &= v_1 + v_2(y) - v_3(x) =\ k_1 + \frac{V_1 y^h}{M_1^h + y^h} - u_1 x \\
\dot y &= v_4(x) - v_5(y) =\ \frac{V_2 x^h}{M_2^h + x^h} - u_2 y,
\end{aligned}
\label{eq:switch-model}
\end{equation}
where $x$ and $y$ denote the amounts of X and Y, respectively.
The vector $(x, y)\T$ will be referred to as the \emph{microstate}
of the biochemical reaction system.
Ultrasensitivity, which is required to achieve bistability \cite{FerrellXio2001},
is generated by the Hill-type production rates $v_2$ and $v_4$.

In the sequel, we will assume that the molecular species X and Y represent gene transcripts,
and the amounts $x$ and $y$ indicate the respective transcript copy number.
The nominal parameter values that we use are given in Table~\ref{tab:param-values}.
For simplicity, we assume that the parameters are symmetric, i.e.\ $V_1 = V_2$,
$M_1 = M_2$ and $u_1 = u_2$.
The parameter values are within the physiological range for typical gene transcription
processes.
In particular, the degradation rate of $0.01 \frac{1}{\mathrm{min}}$ corresponds to
a gene transcript half-life time of about 70 minutes.
Typical transcript half-life times are in a range from several minutes to 
several hours.
The minimal transcription rate of X is given by $k_1$ and corresponds to 3.3 transcripts
that are produced per hour.
The transcription rate upon maximal activation is given by $V_{1,2}$ and corresponds
to 33 transcripts produced per hour.
This is an arguably low transcription rate, but it was mainly chosen to allow for
an efficient computational treatment in the stochastic case.
The basic principle also works with higher maximal transcription rates.

\begin{table}
\caption{Nominal parameter values for the bistable switch model \eqref{eq:switch-model}. 
Transcript copy numbers are considered to be dimensionless.}
\begin{center}
\begin{tabular}{ll|ll}
Parameter & Value  & Parameter & Value \\ \hline
$k_1$ & $0.055\ \frac{1}{\mathrm{min}}$ &  $V_{1,2}$ & $0.55\ \frac{1}{\mathrm{min}}$ \\
$M_{1,2}$ & $25$ &  $h$ & $3$ \\
$u_{1,2}$ & $0.01\ \frac{1}{\mathrm{min}}$ &  &
\end{tabular}
\end{center}
\label{tab:param-values}
\end{table}

\begin{figure}
\begin{center}
\includegraphics{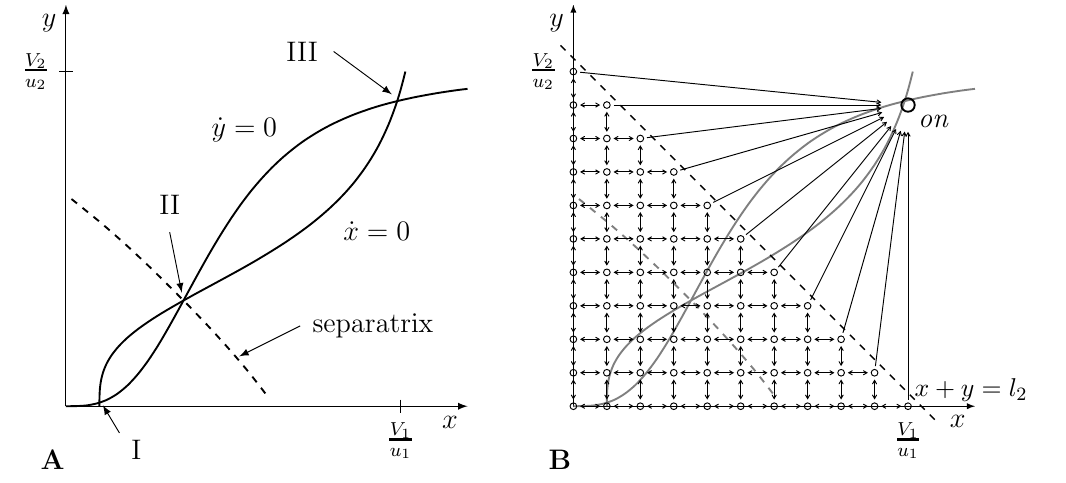}
\end{center}
\caption{Characterisation of the phase space in the bistable switch model \eqref{eq:switch-model}.
\textbf{A}: Nullclines for the deterministic model of the bistable switch.
\emph{I} and \emph{III} are stable equilibrium points, \emph{II} is an unstable one.
\textbf{B}: Schematic illustration of the configuration space for the Markov process \eqref{eq:linear-ode} describing the cell transformation process.}
\label{fig:state-space}
\end{figure}

For two-dimensional systems, it is convenient to check bistability by considering
nullclines in the state space \cite{EissingWal2007}.
With this graphical representation, it is also easy to evaluate how good the
two stable states are actually separated \cite{CherryAdl2000}.
The nullclines for the model given in \eqref{eq:switch-model}, with nominal
parameter values, are depicted in Figure~\ref{fig:state-space}A.
From the figure, it is clear that there are three equilibrium points, labelled
I, II and III.
A stability analysis of the equilibrium points shows that the deterministic system described by \eqref{eq:switch-model} is bistable, and the corresponding reaction network implements a bistable switch.
We construct a \emph{macrostate} for this system by defining the three sets
$\Omega_{off}, \Omega_{on}, \Omega_{trans} \subset \Real^2$, corresponding
to the switch being \emph{off}, \emph{on}, or \emph{in transition}, respectively.
$\Omega_{off}$ contains the equilibrium point I, $\Omega_{trans}$ contains II, 
and $\Omega_{on}$ contains III.
Moreover, to have a well-defined macrostate for each microstate that the system
can reach, we require that $\Omega_{off} \cup \Omega_{on} \cup \Omega_{trans} = \Real_{+}^2$,
where $\Real_{+}$ stands for the non-negative real numbers.
For our model, we define 
\begin{align*}
\Omega_{off} &= \lbrace (x,y) \in \Real_+^2 \mid x + y \leq l_1 \rbrace \\
\Omega_{on} &= \lbrace (x,y) \in \Real_+^2 \mid x + y \geq l_2 \rbrace \\
\intertext{and, consequently,}
\Omega_{trans} &= \lbrace (x,y) \in \Real_+^2 \mid l_1 < x + y < l_2 \rbrace,
\end{align*}
with suitable parameters $l_1$ and $l_2$.
With model parameters as given in Table~\ref{tab:param-values}, a suitable
choice which we will use in this work is $l_1 = 25$ and $l_2 = 55$.

\subsection*{Stochastic model of a bistable switch}
\label{ssec:stoch-switch-model}

The deterministic model of the bistable switch discussed in the previous section is suitable to describe the existence of two distinct macrostates, corresponding to stable equilibrium points in the model.
However, to capture transitions between these macrostates which are caused by intrinsic fluctuations, a stochastic model has to be considered.
The stochastic description of the bistable switch makes use of the Markov property of biochemical reaction networks.
In a stochastic setting, the amounts of molecular species may only take discrete values from the
set $\Omega = \lbrace (x,y)\T \mid x \in \mathbb{N}_0, y \in \mathbb{N}_0 \rbrace$.
The stochastic state of the switch at time $t$ is given by the discrete probability distribution
$P(x,y,t)$, which for each microstate $(x,y) \in \Omega$ gives the probability that
the switch is in this microstate at time $t$. 
To describe the temporal evolution of the probability distribution, we use
the chemical master equation (CME) \cite{Kampen1981}.
The reaction network for the bistable switch is not 
described with elementary reactions only,
and thus it is not possible to construct the CME according to its rigorous
derivation \cite{Gillespie1992}.
However, a theoretical investigation by Rao and Arkin \cite{RaoArk2003} has
shown that as an approximation, the propensity functions for state transitions
can be taken from the according reaction rate laws.
For the bistable switch described above, we thus use the CME
\begin{equation}\begin{aligned}
\dot P(x,y,t) &= - \sum_{i=1}^5 v_i(x,y) P(x,y,t) + v_1 P(x-1,y,t) \\
& + v_2(y) P(x-1,y,t) + v_3(x) P(x+1,y,t) \\
& + v_4(x) P(x,y-1,t) + v_5(y) P(x,y+1,t)
\end{aligned}\end{equation}
for $(x,y)\in\Omega$.

In the stochastic description, we can compute the probabilities that the switch is in
any of its three macrostates directly from a solution of the CME.
Define $p_{off}(t)$, $p_{trans}(t)$ and $p_{on}(t)$ as the probabilities
that the switch is \emph{off}, \emph{in transition} and \emph{on}, respectively.
Given a solution of the CME, these can be computed by summing up the probabilities
that the system is in the corresponding microstates, i.e.\ 
$p_{on}(t) = \sum_{(x,y)\in\Omega_{on}} P(x,y,t)$, and equivalently
for the other macrostates.

\subsection*{A transformation process modelled with a stochastic switch}
\label{ssec:transformation}

Cellular transformation processes are often based on a bistable biochemical or genetic
switch.
In the initial state of the cell, the switch would be in the \emph{off} state.
Switching to the \emph{on} state implies a significant change in the amount of
an involved signaling molecule, e.g.\ a transcription factor.
If the \emph{on} state is maintained for some time, this change would result
in a larger phenotypical change of the cell, e.g.\ through significant
changes in gene expression.
The mechanisms that induce this change are not part of the stochastic
switch, but from a signaling perspective downstream of it.

Most transformation processes rely on specific external stimuli, and the cell
will initiate the transformation upon encountering the required stimulus.
There are however examples where such a stimulus is not strictly required,
and this is the case that we are dealing with in this paper.
Moreover, we will focus on the behavior of cell populations, studying the problem
how the temporal dynamics of the transformation process evolve
in a pool of many cells.

The basic mechanism that actually triggers the bistable switch in our model without
an external stimulus are the intrinsic fluctuations of concentrations
in any biochemical reaction network, that are due to the stochastic nature of
chemical reactions.
As a rare event, these fluctuations may become so large that the
microstate of the system crosses the separatrix between the domains of attraction
in the deterministic system.
As a consequence, the microstate around the other stable equilibrium point will become strongly
attractive, and the switch will change its macrostate to \emph{on} with a high probability.
In this paper, we assume that the transformation is irreversible, which fits well to the process
of follicle growth initiation.
Also other processes such as programmed cell death are irreversible.

The described transformation process is easily modelled as a continuous-time
Markov process.
If the switch is in one of the macrostates \emph{off} or \emph{in transition}, then
we directly use the microstates and transition probabilities of the underlying biochemical
reaction network to model the transformation process.
To account for the irreversibility of the transformation, the \emph{on} state of the switch
corresponds to only one state of the Markov process, which is an absorbing
state.
The transitions of other microstates to the absorbing state are governed by the
propensity functions for the corresponding transitions in the underlying biochemical network.
The resulting state space for the Markov process model of the transformation
process is shown in Figure~\ref{fig:state-space}B.

In our model of the stochastic switch, the macrostates \emph{off} and \emph{in transition}
are defined by compact regions in state space.
As a consequence, the Markov model of the considered transformation process
has a finite state space, and can therefore be treated computationally with standard approaches.
Let $\mathbf{P}(\mathbf{x},\mathbf{y},t)\in\Real^n$ denote the complete probability state vector of the system
with $n$ discrete states $(\mathbf{x},\mathbf{y})$,
\begin{equation}
\begin{aligned}
\mathbf{P}(\mathbf{x},\mathbf{y},t) = &\ (P(0,0,t),\ P(1,0,t),\ P(0,1,t),\ \ldots,\ P(\frac{V_1}{u_1},0,t),\ P(x_c,y_c,t))\T,
\end{aligned}
\end{equation} 
where $(x_c,y_c)$ stands for the microstates which correspond to the \emph{on} state of the switch (see Figure~\ref{fig:state-space}B).
The \emph{on} state is made irreversible by removing the associated transitions from the Markov process.
The master equation can be written as the linear ordinary differential equation
\begin{equation}
\label{eq:linear-ode}
\dot{\mathbf{P}}(\mathbf{x},\mathbf{y},t) = A\,\mathbf{P}(\mathbf{x},\mathbf{y},t),
\end{equation}
where $A \in \Real^{n\times n}$ is the state transition matrix.
The matrix $A$ can be computed directly from the values of the reaction propensity functions in
each microstate \cite{MunskyKha2006}.

The differential equation \eqref{eq:linear-ode} can be solved using standard tools
for numerical integration.
For the results described in this paper, we used the \texttt{ode15s} function in \textsc{Matlab} (The MathWorks, Natick, MA) to obtain a numerical solution of \eqref{eq:linear-ode}.

\section*{Results and Discussion}
\label{sec:results}

\subsection*{A hypothetical mechanism for oocyte maturation}

In this section, we suggest a biochemical mechanism that offers a molecular explanation
for the large depletion times of several decades in the human oocyte pool.
The model is based on experimental evidence obtained in a very informative
series of studies by Skinner and colleagues (see \cite{Skinner2005} for a review), where
the influence of several ovarian factors on the primordial to primary
transition as well as some interactions between them have been studied.
Because a positive feedback loop is necessary for a bistable switch \cite{KaufmanSou2007},
we have specifically searched for such an interconnection.

Indeed, experimental evidence suggests a positive feedback circuit involving
two ovarian factors that are relevant in the primordial to primary transition:
the factor KIT ligand (KITL) is produced by granulosa cells surrounding the oocyte
and stimulates both the oocyte and surrounding theca cells to promote follicle development.
Moreover, KITL stimulates the production of both keratinocyte growth factor (KGF) and
hepatocyte growth factor (HGF) in
the surrounding theca cells.
KGF and HGF themself stimulate the production of KITL in the granulosa cells,
thus providing a positive feedback loop \cite{ParrottSki1998}.
Moreover, the oocyte of primordial and developing follicles produces basic fibroblast
growth factor (bFGF), which acts on surrounding granulosa cells and has been
shown to increase the expression of KITL \cite{NilssonSki2004}.

\begin{figure}
\begin{center}
\includegraphics{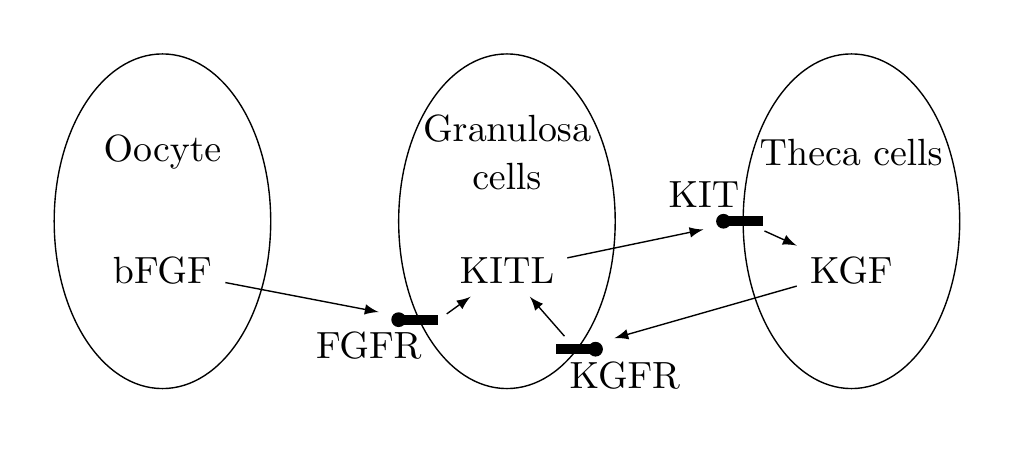}
\end{center}
\caption{Hypothetical biochemical network for the primordial to primary transition in ovarian follicles}
\label{fig:oocyte-model}
\end{figure}

These pieces of experimental evidence thus support the hypothetical mechanism
that is shown in Figure~\ref{fig:oocyte-model}.
Our simplistic mathematical model presented in \eqref{eq:switch-model} can be used to describe this
mechanism, where the variable $x$ represents granulosa-derived KITL activity and 
$y$ represents theca-derived KGF and HGF activity.
The reaction $v_1$ describes the influence on KITL expression of oocyte-derived bFGF, which
is here assumed to be constant.
The reactions $v_2$ and $v_4$ arise from the positive feedback interconnection,
whereas $v_3$ and $v_5$ describe a constitutive degradation of KITL, KGF and HGF.

\subsection*{The stochastic switch generates reliable long-term behavior}
\label{ssec:long-term-behavior}

The differential equation \eqref{eq:linear-ode} that governs the initiation probabilities
of the irreversible transformation process is a linear ordinary differential equation,
so in principle it can be solved analytically.
Due to the size of the system, this is however not feasible.
Yet, we can characterize the probability that a given cell has
initiated the transformation process by the explicit formula
\begin{equation}\begin{aligned}
p_{on}(t) = 1 - c_1 e^{-\lambda_1 t} + \sum_{i=2}^n c_i(t) e^{-\lambda_i t},
\end{aligned}
\label{eq:analytical-solution}
\end{equation}
where $c_1 \geq 0$, $0 < \lambda_1 < \mathrm{Re}(\lambda_i)$ for $i=2,\ldots,n$, and the
$c_i(t)$ are polynomials in $t$.
The mathematical derivation of \eqref{eq:analytical-solution} is
provided in the appendix.

\begin{figure}
\begin{center}
\includegraphics{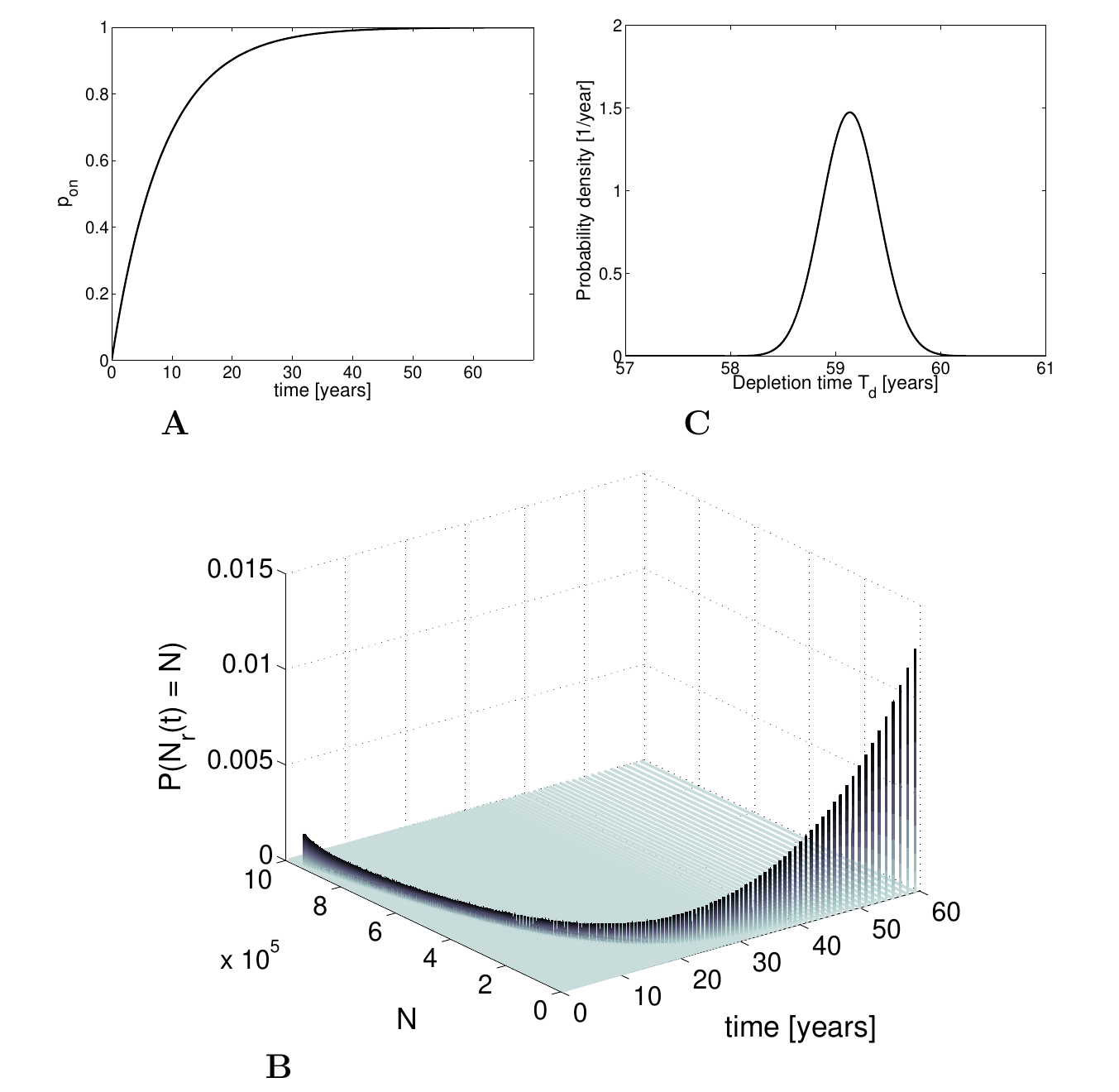}
\end{center}
\caption{Dynamical characteristics of the stochastic bistable switch on the single cell level and the population level.
\textbf{A}: Probability of transformation event $p_{on}(t)$.
\textbf{B}: Population size probability distribution over time.
\textbf{C}: Probability density function of the depletion time $T_d$.}
\label{fig:dynamics}
\end{figure}

From considering the general form of the analytical solution given in \eqref{eq:analytical-solution}, we
obtain two important conclusions about the stochastic transformation process.
First, we observe that the probability for a given cell to initiate the transformation
tends to 1 as time increases.
Second, because $\lambda_1$ is the dominant decay rate, for larger times $t \gg 0$ the probability of 
not having initiated the transformation can be approximated by
$1 - p_{on}(t) \approx c_1 e^{-\lambda_1 t}$, a simple exponential decay.
For the biochemical parameter values given in Table~\ref{tab:param-values}, the
numerical solution for $p_{on}(t)$ is shown in Figure~\ref{fig:dynamics}A.
For these parameter values, which are in the physiological range for the
considered biological processes, we indeed get to a time scale of years to decades
in the probability of the transformation event,
with a half-life time of about $5.9$ years.

Let us now move to the population level, and consider a pool of cells, each
of them being subject to the considered transformation process with a bistable
switch.
In the first step, we make the simplistic assumption that no interactions among the cells are taking place, so individual
transformations are probabilistically independent events.
The number of remaining cells $N_r(t)$ can easily be
characterized by a binomial distribution as
\begin{equation}\begin{aligned}
\label{eq:total-distribution}
P(N_r(t)=N) = {N_0 \choose N} (1-p_{on}(t))^N p_{on}(t)^{N_0 - N}.
\end{aligned}\end{equation}
The properties of the binomial distribution give the expected number of cells
remaining in the pool as
\begin{equation}\begin{aligned}
E[N_r(t)] = N_0  (1-p_{on}(t)),
\end{aligned}\end{equation}
where $N_0$ is the initial number of cells in the pool.

The probability distribution $P(N_r(t)=N)$ for the population size in the transformation process considered in this paper 
is shown in Figure~\ref{fig:dynamics}B as a function of both cell number $N$ and time $t$.
The number of initial cells $N_0 = 10^6$ was chosen from the reported
range of ovarian follicles, $7\cdot10^5$ to $2\cdot10^6$, in human females at birth.
For each point in time, the distribution has a very sharp peak, which indicates that the average value 
$E[N_r(t)]$ is a reliable prediction for the number of cells that have already
undergone the transformation at a given time.

A relevant characteristic of the considered process is the time at which
the initial cell population is depleted, i.e.\ when nearly all cells have
undergone the transformation.
To make this notion precise, we introduce the depletion number $N_d$.
The depletion time $T_d$ is defined as the smallest time $t$ such that $N_r(t) \leq N_d$,
i.e.\ only $N_d$ cells are remaining in the initial population.
For the process of follicle growth initiation, we use $N_d = 1,000$, which
has been considered to mark the onset of menopause \cite{FaddyGos1996}.

The cumulative probability distribution function for the depletion time $T_d$
is computed from the distribution obtained in \eqref{eq:total-distribution} as 
\begin{equation}
\begin{aligned}
\label{eq:td-distribution}
P(T_d \leq t) = P(N_r(t) \leq N_d) = \sum_{N\leq N_0} P(N_r(t) = N).
\end{aligned}
\end{equation}
The probability density function for the depletion time is computed by taking the derivative
of the cumulative probability distribution function \eqref{eq:td-distribution}.
The resulting probability density function for the depletion time
in follicle growth initiation is shown in Figure~\ref{fig:dynamics}C.
From the density function, the expected value and the standard deviation are obtained as
$E[T_d] = 59.1$ years and $\sqrt{E[T_d^2]-E[T_d]^2} = 0.27$ years, respectively.

The expected value for $T_d$ can also be computed by solving $1-p_{on}(T_d) = \frac{N_d}{N_0}$.
Using \eqref{eq:analytical-solution}, it can thus be approximated by
$E[T_d] \approx \frac{1}{-\lambda_1} \ln \frac{N_d}{N_0}$, where $\lambda_1$ is the
dominant decay rate of the process.

Next, we compare the computed statistical characteristics of the follicle depletion process
to medical data.
Explicit follicle counts are only sparsely available.
However, the available pieces of data indicate that fluctuations in 
actual follicle numbers are larger than predicted by our model \cite{GougeonEco1994}.
Concerning the depletion time, a recent medical study suggest an average 
age of $51.1$ years for the onset of menopause, with a standard deviation of $3.8$ years \cite{BruinBov2001}.
Our model predicts a depletion time of $T_d = 59.1$ years, which is reasonably close to the experimentally observed depletion time.
However, the standard deviation of 0.27 years in our model is significantly less than observed from
medical data.
In summary, even though our model is based on a highly stochastic process, the analysis reveals
that it leads to much more reliable temporal characteristics than observed in the real system.
This indicates that stochastic effects alone may not be sufficient to explain the heterogeneity observed in the follicle depletion process.

An alternative explanation would be by heterogeneous parameter values among individual organisms.
This explanation is also supported by statistical analyses of medical data \cite{BruinBov2001}, where it is suggested
that the onset of menopause is largely based on genetic factors,
which would be related to parameter values in our model.
To investigate this possibility, we have computed the expected depletion times for different
parameter values.
The computation was based on the eigenvalues of the transition matrix $A$ and
the approximation $E[T_d] \approx \frac{1}{-\lambda_1} \ln \frac{N_d}{N_0}$.
The results are given in Table~\ref{tab:depletion-times}.
From these results, we note that even small parameter variations in the model of the bistable switch lead to very large variations in the expected depletion time.
This is not realistic for a biological system, and in the following section we explore mechanisms to increase the robustness of the depletion time with respect to parameter variations.

\begin{table}
\caption{Expected depletion times (years) in the single cell model \eqref{eq:linear-ode}.
Expected depletion times (years) in the single cell model \eqref{eq:linear-ode} for multiplicative variations in single parameters.
For simplicity, we always assume $V_1=V_2$, $u_1 = u_2$, and $M_1 = M_2$.
}
\begin{center}
\begin{tabular}{l|cccccc}
\raisebox{-0.5ex}{Param.}\hspace*{1.5ex}\raisebox{0.5ex}{Factor} & 0.8 & 0.9 & 0.95 & 1.05 & 1.1 & 1.3 \\ \hline
$k_1$ & 1300 & 254 & 120 & 31 & 16 & 1.9 \\
$V_{1,2}$ & 1720 & 322 & 135 & 29 & 15 & 2.2 \\
$u_{1,2}$ & 0.4 & 4.1 & 15 & 248 & 1010 & $1.2\cdot10^5$ \\
$M_{1,2}$ & 0.2 & 2.1 & 10 & 418 & 3410 & $2.0\cdot10^7$ \\
$h$ & 0.6 & 6.3 & 20 & 163 & 420 & $9.6\cdot10^3$
\end{tabular}
\end{center}
\label{tab:depletion-times}
\end{table} 

\subsection*{Increased robustness by interactions on the population level}

In the last section, we have characterized the properties of the transformation
process based on a bistable switch, with the depletion time of a pool of cells
subject to the transformation as characteristic output of the model.
We have shown that the proposed model produces reliable depletion times,
in the sense of a small standard deviation, for fixed values
of the biochemical parameters.
However, we have also observed that the average depletion time in the basic model is quite sensitive
to variations in the biochemical parameters.
Clearly, this large sensitivity is not acceptable in a model that should be a meaningful representation of the primordial to primary follicle transition.
In this section, we propose an additional mechanism that reduces
the sensitivity of the average depletion time significantly.

The additional mechanism is based on the experimental observation that
follicles in later stages of development actively suppress the primordial
to primary transition in resting follicles \cite{Skinner2005}.
The inhibition of follicle growth initiation is mediated by the Anti-M\"ullerian hormone (AMH),
which interferes with stimulatory signals by KITL, bFGF, and KGF \cite{NilssonRog2007}.
Although AMH is known to signal via SMAD proteins \cite{VisserThe2005}, the molecular mechanisms of follicle growth inhibition by AMH seem to be unknown.
To include the inhibitory effect into the simplistic switch model \eqref{eq:switch-model}, we assume that the rate of KITL production in primordial follicles is reduced with an increasing number of growing follicles.
This is achieved by changing $k_1$ in the original model given in \eqref{eq:switch-model} from a constant parameter to the expression
\begin{equation}
\begin{aligned}
k_1(n_2) = \frac{k_{1,max} K_n}{K_n + n_2},
\end{aligned}
\end{equation}
where $k_{1,max}$ is the maximal production rate of KITL, $n_2$ is the number of growing follicles, and $K_n$ is an additional parameter.
While follicle development is a complex process with many intermediate stages \cite{YehAda1999}, in this analysis we use a simple two-state population model, where $n_1$ denotes the number of primordial follicles, and $n_2$ the number of growing follicles.
The assumptions of the model are that primordial follicles initiate growth with a rate as determined by $\lambda_1$ in \eqref{eq:analytical-solution}.
Growing follicles are assumed to stay in this stage for a constant amount of time $\tau$, after which they leave the pool either through ovulation or atresia.
From these specifications, one can derive a model given by the system of delay-differential equations
\begin{equation}
\label{eq:pop-model}
\begin{aligned}
\dot n_1(t) &= - \lambda_1 (n_2(t)) n_1(t) \\
\dot n_2(t) &= \lambda_1(n_2(t)) n_1 (t) - \lambda_1(n_2(t-\tau)) n_1(t-\tau),
\end{aligned}
\end{equation}
where $\lambda_1(n_2)$ is the decay rate computed from the transition matrix $A$ in \eqref{eq:linear-ode}.

\begin{figure}
\begin{center}
\includegraphics[width=8cm]{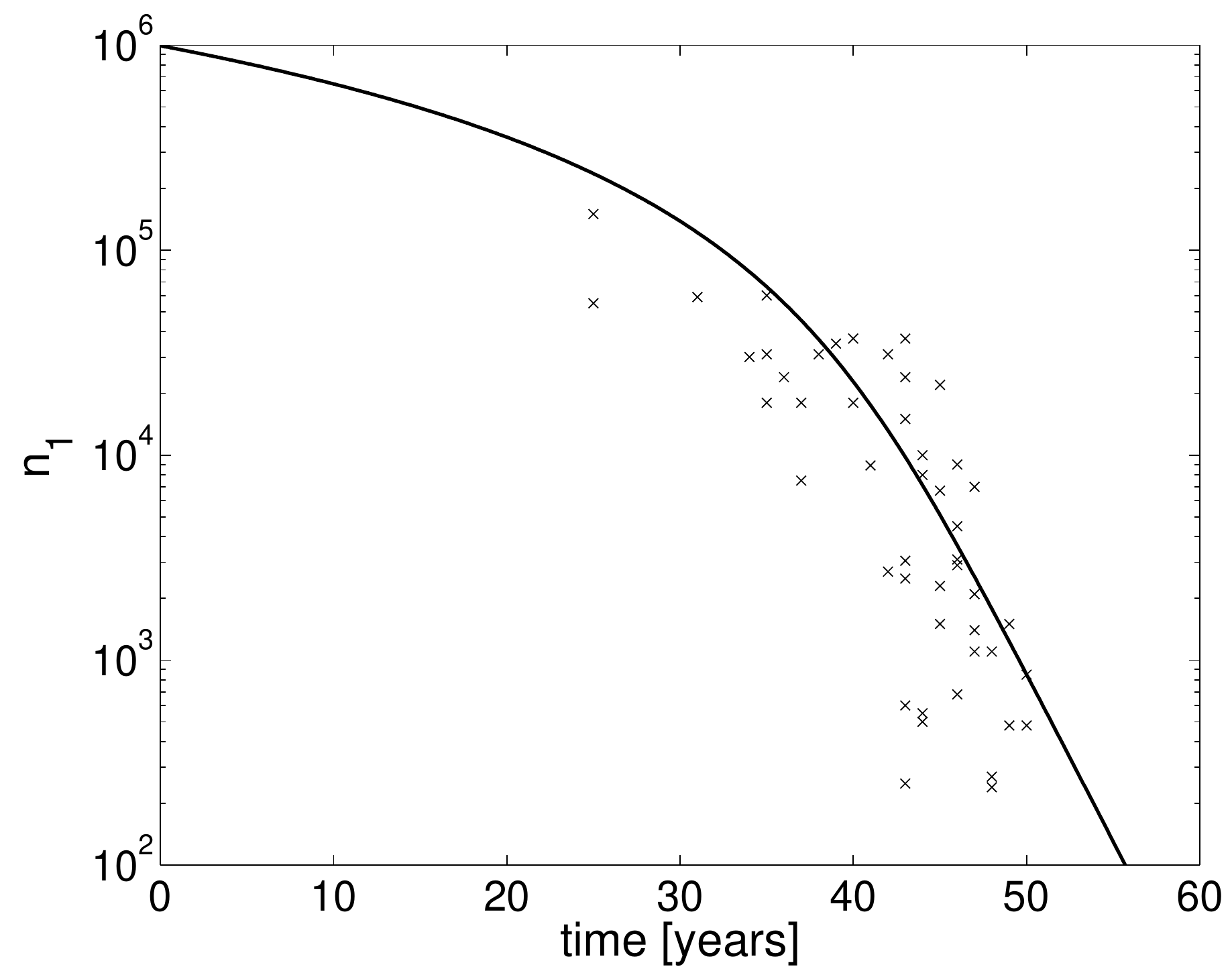}
\end{center}
\caption{Evolution of follicle number.
Model predictions from \eqref{eq:pop-model} (\emph{line}) vs.\ clinical data from \cite{FaddyGos1995} (\emph{crosses}).}
\label{fig:pop-trajectory}
\end{figure}

Using the parameters in Table~\ref{tab:param-values2}, the population model given by \eqref{eq:pop-model} now predicts a depletion time of $T_d = 50.0$ years, which is almost equal to the depletion time suggested by the medical study \cite{BruinBov2001}.
The development of the ovarian follicle pool over time, as predicted by the model in \eqref{eq:pop-model}, is shown in Figure~\ref{fig:pop-trajectory}.
The prediction is compared to clinical data of follicle numbers at different ages taken from \cite{FaddyGos1995}.
Although the parameters have only been adjusted to fit the depletion time, the predicted time course is reasonable close to the clinical data.
In particular, the proposed population model \eqref{eq:pop-model} intrinsically captures the previously observed increase in the follicle depletion rate at an age of approximatively 38 years \cite{FaddyGos1995}.

\begin{table}
\caption{Nominal parameter values for the population model \eqref{eq:pop-model}.}
\begin{center}
\begin{tabular}{ll|ll}
Parameter & Value  & Parameter & Value \\ \hline
$k_{1,max}$ & $0.06\ \frac{1}{\mathrm{min}}$ &  $V_{1,2}$ & $0.55\ \frac{1}{\mathrm{min}}$ \\
$M_{1,2}$ & $25$ &  $h$ & $3$ \\
$u_{1,2}$ & $0.01\ \frac{1}{\mathrm{min}}$ &  $K_n$ & $8\cdot10^4$ \\
$\tau$ & 0.4 years & &
\end{tabular}
\end{center}
\label{tab:param-values2}
\end{table} 

In order to investigate the sensitivity of the extended model to variations in the biochemical parameters, we have computed again the expected depletion times for different parameter values.
The results are given in Table~\ref{tab:depletion-times2}.
The variation in the depletion time is significantly reduced, compared to the model \eqref{eq:linear-ode}, where follicle interactions are neglected.
It should also be pointed out that the depletion time is quite insensitive towards variations in the two parameters $K_n$ and $\tau$ which were newly introduced in the population model.
This result illustrates that the robustness of the depletion time with respect to parameter variations may be substantially increased by adding interactions among individual follicles to the proposed model of the transformation process.

\begin{table}
\caption{Expected depletion times (years) in the population model \eqref{eq:pop-model} for multiplicative variations in single parameters.}
\begin{center}
\begin{tabular}{l|cccccc}
\raisebox{-0.5ex}{Param.}\hspace*{1.5ex}\raisebox{0.5ex}{Factor} & 0.8 & 0.9 & 0.95 & 1.05 & 1.1 & 1.3 \\ \hline
$k_1$ & 450 & 120 & 74 & 37 & 29 & 15 \\
$V_{1,2}$ & $>500$ & 120 & 78 & 36 & 27 & 14 \\
$u_{1,2}$ & 9.4 & 18 & 28 & 120 & 330 & $>500$ \\
$M_{1,2}$ & 6.8 & 14 & 24 & 160 & $>500$ & $>500$ \\
$h$ & 9.5 & 20 & 31 & 88 & 160 & $>500$ \\
$K_n$ & 56 & 53 & 52 & 49 & 48 & 44 \\
$\tau$ & 45 & 48 & 49 & 51 & 52 & 57 \\
\end{tabular} 
\end{center}
\label{tab:depletion-times2}
\end{table} 

\section*{Conclusions}
\label{sec:conclusions}

In this paper, we deal with a fundamental question in the development of 
multicellular organisms: How can biochemical reactions and genetic mechanism acting on the scale
of minutes in individual cells generate dynamics with characteristic times
of years to decades on the tissue level?
As a possible mechanism to achieve this, we propose a generic transformation process
based on a bistable stochastic switch.
From the underlying genetic interactions and biochemical reactions, the process can be modelled as a continuous-time Markov process.
We show that the proposed stochastic mechanism generates reliable long-term behavior on the population level, with cells undergoing the transformation with an exponentially decaying rate.
Thereby, the decay rate is equal to the dominant eigenvalue of the transition matrix describing the underlying biochemical network.

We pose the hypothesis that a biological instance of this mechanism is present in the development of ovarian follicles.
To describe this process, we constructed a simple model of a bistable switch in the primordial to primary transition for ovarian follicles.
The model is based on experimentally determined factors and their interactions in the different cell types constituting the ovarian follicles.
Although it is not assured that a bistable switch in ovarian follicles will indeed be based on the factors that we have used here, the basic mechanism would work equivalently well with other factors.

Despite its simplicity, our model explains well how the long-term characteristics of follicle development may reliably be generated by biochemical reactions occurring on much shorter time scales.
Keeping the model simple serves two purposes: first, it shows that the dynamics of follicle growth initiation can be generated by a quite simple mechanism.
Clearly, additional pathways and regulatory feedback interactions that we have not included in this model can be expected to be present in the system.
These may serve to increase robustness of the network, or to provide additional inputs to control the transition rate, e.g.\ for the endocrine system.
Second, the simplicity of the model allows us to solve the chemical master equation for the network numerically, and thus to obtain a quantitative description of the model behavior.

As a possible shortcoming of the basic model on the single cell level, we observe an unrealistic large sensitivity of the follicle depletion time with respect to parameter variations.
By adding the experimentally supported inhibition of follicle growth initiation by later-stage growing follicles, the sensitivity of the depletion time could be reduced significantly.
Apart from the inhibition included in the model, other interactions among individual follicles seem to play a role in the primordial to primary transition \cite{Silva-ButtkusMar2009}.
We envision that the inclusion of more regulatory interactions may further decrease the sensitivity of the depletion time with respect to parameter variations to a physiologically plausible level.

\section*{Appendix: Computation of the transition probability}

In this section, we prove that the probability that a given cell has undergone the considered transformation process is given by $p_{on}(t)$ as in \eqref{eq:analytical-solution}. The proof is based on considering the solution of the underlying CME \eqref{eq:linear-ode}.

Since the last microstate is an absorbing state of the Markov process, \eqref{eq:linear-ode} can be written as
\begin{equation}
\begin{aligned}
\dot{\mathbf P} &= \begin{pmatrix}A_{red} & 0 \\ a_{abs} & 0\end{pmatrix} \mathbf P,
\end{aligned}
\end{equation}
where $A_{red} \in \Real^{(n-1) \times (n-1)}$ describes the interactions among the non-absorbing states, and $a_{abs} \in \Real^{1 \times (n-1)}$ describes the transition propensities to the absorbing state.

Let us first derive some essential properties of the matrix $A_{red}$.
Since $A$ is a stochastic matrix, we have 
\begin{equation}
\label{eq:diagonally-dominant}
\begin{aligned}
\sum_{j=1,\ j\neq i}^{n-1} \vert A_{ji} \vert \leq - A_{ii},
\end{aligned}
\end{equation}
for $i = 1,\dotsc,n-1$, i.e.\ $A_{red}$ is diagonally dominant.
Thus, Gersgorin's theorem \cite{LancasterTis1985} asserts that all eigenvalues of $A_{red}$ have a non-positive real part.
Even more, since $a_{abs}$ is non-zero, \eqref{eq:diagonally-dominant} holds with a strict inequality for at least one $i$.
Thus, by Theorem~10.7.2 in \cite{LancasterTis1985}, all eigenvalues of $A_{red}$ have negative real part.
By the properties of the considered biochemical network, $A_{red}$ is irreducible, and its off-diagonal elements are non-negative.
From Corollary~4.3.2 in \cite{Smith1995}, it follows that $A_{red}$ has an eigenvalue $\lambda_1 \in \Real$ with algebraic multiplicity 1 and a strictly positive corresponding eigenvector $v_1$ such that $\mathrm{Re}\,\lambda < \lambda_1$ for all $\lambda \neq \lambda_1$ in the spectrum of $A_{red}$.

Denoting $\mathbf P_{red} = (P_1, \dotsc, P_{n-1})\T$, we have $\dot{\mathbf P}_{red} = A_{red} \mathbf P_{red}$.
From the previously derived properties of the matrix $A_{red}$, the general solution of this differential equation is given by
\begin{equation}
\begin{aligned}
\mathbf P_{red}(t) &= \tilde a v_1 e^{\lambda_1 t} + \sum_{i=2}^s v_i \tilde c_i(t) e^{\lambda_i t},
\end{aligned}
\end{equation}
where $\tilde c_i(t)$ are polynomials in $t$ and $\tilde a$ is a constant coefficient, depending on the initial condition $\mathbf P_{red}(0)$.
The condition $P_{red}(t) \geq 0$ for all $t$ implies that $\tilde a \geq 0$.
The transition probability $p_{on}(t)$ is computed as
\begin{equation}
\begin{aligned}
p_{on}(t) &= 1 - \mathbf 1\T P_{red}(t) \\
&= 1 - a e^{\lambda_1 t} + \sum_{i=2}^s c_i(t) e^{\lambda_i t},
\end{aligned}
\end{equation}
where $a = \tilde a \mathbf 1\T v_1 \geq 0$ and $c_i(t) = \tilde c_i(t) \mathbf 1\T v_i$,
$\mathbf 1 = (1,\dotsc,1)\T$.
    
%

\section*{Acknowledgements}
SW and FA acknowledge support by the German Research Foundation (DFG) through the \textit{Cluster of Excellence in Simulation Technology} (EXC 310) at the University of Stuttgart.

\end{document}